\documentclass[aps,prl,superscriptaddress,twocolumn,showpacs]{revtex4}

\usepackage{bm}
\usepackage{graphicx}
\usepackage{amsfonts}
\usepackage{amsmath}

\begin{document}

\title{Studying quantum spin systems through entanglement estimators}

\author{Tommaso Roscilde}
\affiliation{Department of Physics and Astronomy, University of 
Southern California, Los Angeles, CA 90089-0484}
\author{Paola Verrucchi}
\affiliation{\mbox{Istituto Nazionale per la Fisica della Materia, UdR Firenze,
Via G. Sansone 1, I-50019 Sesto F.no (FI), Italy}}
\author{Andrea Fubini}
\affiliation{\mbox{Istituto Nazionale per la Fisica della Materia, UdR Firenze,
Via G. Sansone 1, I-50019 Sesto F.no (FI), Italy}}
\affiliation{\mbox{Dipartimento di Fisica dell'Universit\`a di Firenze,
Via G. Sansone 1, I-50019 Sesto F.no (FI), Italy}}  
\author{Stephan Haas}
\affiliation{Department of Physics and Astronomy, University of 
Southern California, Los Angeles, CA 90089-0484}    
\author{Valerio Tognetti}
\affiliation{\mbox{Istituto Nazionale per la Fisica della Materia, UdR Firenze,
Via G. Sansone 1, I-50019 Sesto F.no (FI), Italy}}
\affiliation{\mbox{Dipartimento di Fisica dell'Universit\`a di Firenze,
Via G. Sansone 1, I-50019 Sesto F.no (FI), Italy}}  
\affiliation{\mbox{Istituto Nazionale di Fisica Nucleare, Sez.
di Firenze, Via G. Sansone 1, I-50019 Sesto F.no 
(FI), Italy}}

\date{\today}

\begin{abstract}

We study the field dependence of 
the entanglement of formation in anisotropic 
$S=1/2$ antiferromagnetic chains displaying a 
$T=0$ field-driven quantum phase transition.
The analysis is carried out via 
Quantum Monte Carlo simulations.
At zero temperature the entanglement estimators show
abrupt changes at and around criticality, 
vanishing below the critical field, 
in correspondence with an exactly factorized state,
and then immediately recovering a finite value upon
passing through the quantum phase transition.
At the quantum critical point, a deep minimum in the 
pairwise-to-global entanglement ratio shows
that {\it multi-spin entanglement} is strongly enhanced;
moreover this signature represents a novel way of detecting 
the quantum phase transition of the system, relying entirely
on entanglement estimators.


\end{abstract}

\maketitle

 Collective behavior in many-body 
quantum systems is associated with the developement 
of classical correlations, as well as 
of correlations which cannot be accounted for in terms
of classical physics, namely entanglement. Entanglement 
represents in essence the impossibility of giving a 
{\it local} description of a many-body quantum state. 
In particular entanglement is expected to play an essential 
role at quantum phase transitions, where quantum fluctuations 
manifest themselves at all length scales. 
The behavior of entanglement at quantum phase transitions
is a very recent topic, so far 
investigated in a few exactly solvable cases 
\cite{OsborneN02,Osterlohetal02,
Vidaletal03, Verstraeteetal04}.
Moreover, entanglement overwhelmingly comes into play
in quantum computation and communication theory,
being the main physical {\it resource} needed for their
specific tasks \cite{NielsenC00}.
In this respect, the perspective of manipulating
entanglement by tunable quantum many-body effects
appears very intriguing. 

In this letter we show that entaglement estimators give important 
insight in the physics of spin systems. In particular, we focus on 
two striking features
of anisotropic spin chains in an external field: the occurrence of a 
factorized ground state at a
field $h_{\rm f}$ and of a quantum phase transition at $h_{\rm c}$. 
We propose a novel estimator to understand the role of quantum 
fluctuations in the quantum critical region.

We focus our attention on the 1D {\it XYZ model} in a 
field:
\begin{equation}
{\hat{\cal H}} =
- J \sum_i \Big[ \hat{S}^x_i\hat{S}^x_{i+1}
+ \Delta_y \hat{S}^y_i\hat{S}^y_{i+1} 
 - \Delta_z \hat{S}^z_i\hat{S}^z_{i+1}+ h\hat{S}^z_i \Big]
\label{e.XYZhz}
\end{equation}
where $J>0$ is the exchange coupling, $i$ 
runs over the sites of the chain, and 
$h\equiv g\mu_{\rm B} H/J$ is the reduced magnetic field.
In Eq. (\ref{e.XYZhz}) we have implicitly performed 
the canonical transformation 
$\hat{S}^{x,y}_i \to (-1)^{i} \hat{S}^{x,y}_i$ with respect
to the more standard antiferromagnetic hamiltonian.  
The parameters $\Delta_y, \Delta_z \geq 0 $ control the anisotropy 
of the system. In particular, for $\Delta_z = 0$ Eq.~(\ref{e.XYZhz})
reduces to the exactly solvable XY model in a transverse
field \cite{BarouchMD71}. For $\Delta_z \neq 0$
the model does not admit an exact solution
\cite{KurmannTM82}, and it has 
been recently studied within approximate analytical and numerical
approaches \cite{Dmitrievetal02,CapraroG02,Cauxetal03}.
Interestingly, the general model with finite $\Delta_z$
is experimentally realized by the $S=1/2$ quantum spin 
chain Cs$_2$CoCl$_4$ \cite{Kenzelmannetal02}, with strong
planar XZ anisotropy, $\Delta_y \approx 0.25$, 
$\Delta_z \approx 1$, and $J\approx 0.23$ meV.

 In our study, we concentrate on the case
$0 \leq \Delta_y \leq 1$, 
$\Delta_z = 1$, defining the {\it XYX model} 
in a field \cite{DelicaL90}. The more general case of the XYZ model
in a magnetic field with $\Delta_z < 1$ 
should exhibit the same qualitative behavior, as it 
shares the same symmetries with the XYX model in a field, 
and therefore it belongs to the same universality class. 
The analysis is carried out via
Stochastic Series Expansion (SSE) Quantum
 Monte Carlo (QMC) simulations, based on a modified directed-loop
algorithm \cite{SyljuasenS02,note}, for chains of
various length, from $L=40$ to $L=120$.
Ground state properties have been determined by considering
inverse temperatures $\beta=2L$, in order to capture the $T=0$
behavior.

\begin{figure}
\null\vskip -2cm
\includegraphics[
bbllx=100pt,bblly=60pt,bburx=450pt,bbury=500pt,%
     width=40mm,angle=0]{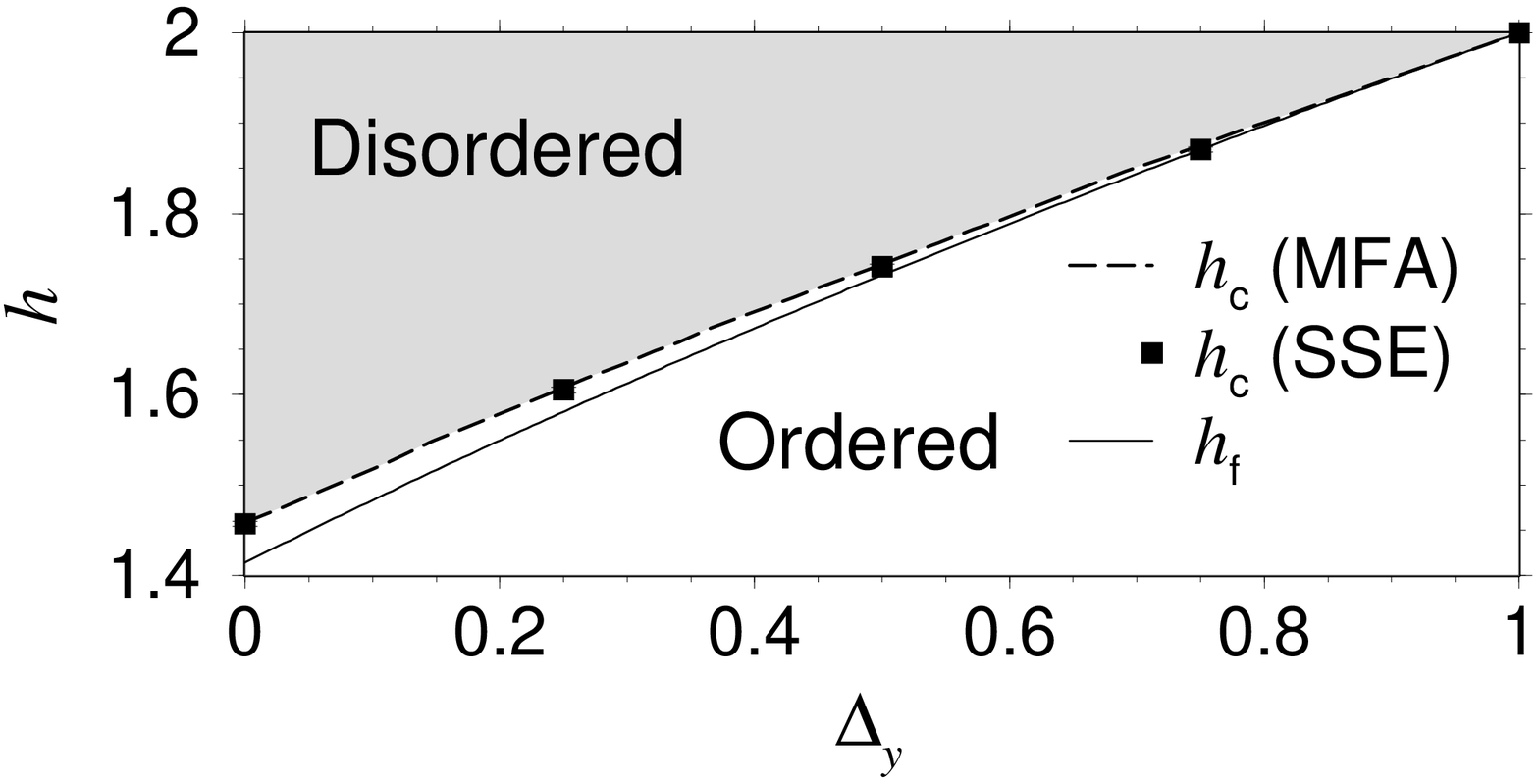} \\
 \includegraphics[
bbllx=100pt,bblly=60pt,bburx=450pt,bbury=500pt,%
     width=40mm,angle=0]{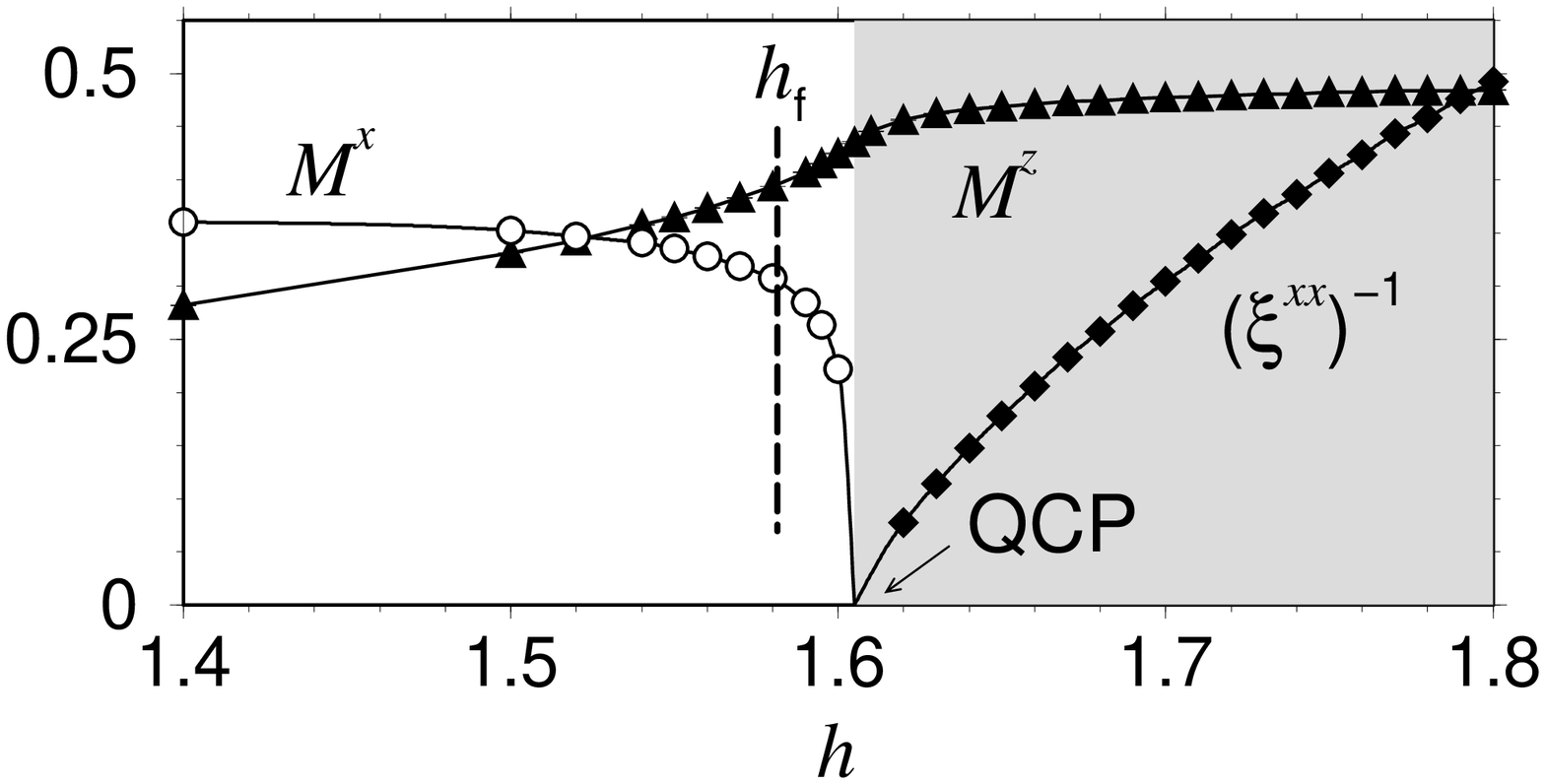}    
\null\vskip -2cm 
\caption{\label{f.XYXphd} Upper panel: ground state phase 
 diagram of the
 XYX model in a field. Mean-field (MFA) results are taken 
 from  Ref. \onlinecite{Cauxetal03}, the factorizing field
 $h_{\rm f}$ from Ref. \onlinecite{KurmannTM82}. 
 Lower panel: quantum critical
 behavior of $x-$ and $z-$magnetizations and 
 correlation length for
 the model with $\Delta_y=0.25$, $L=100$, $\beta=200$.
 The factorizing field is indicated by a dashed line.
 The arrow indicates the quantum critical point (QCP).}
 \null\vskip -0.7cm
\end{figure}

 The ground-state phase diagram of the XYX model in
the $\Delta_y-h$ plane is shown in Fig.~\ref{f.XYXphd}.
The model displays a field-driven quantum phase transition at a
critical field $h_{\rm c}(\Delta_y)$, which separates the
N\'eel-ordered phase ($h\leq h_{\rm c}$)
 from a disordered phase ($h>h_{\rm c}$) with short-range 
 antiferromagnetic correlations
 \cite{KurmannTM82,Dmitrievetal02,Cauxetal03}.
 The transition line $h_{\rm c}(\Delta_y)$ has been determined by 
 a scaling analysis of the correlation length $\xi^{xx}$,
 whose linear scaling $\xi^{xx}\sim L$ marks the quantum critical
 point. Using the critical scaling of the structure factor $S_{xx}(q = 0) \sim
L^{\gamma/\nu - z}$, we verified that the transition belongs to the 1D
transverse-field Ising universality class ($\gamma/\nu=7/4$
and $z=1$), in agreement with analytical predictions \cite{Dmitrievetal02}.

 Besides its quantum critical behavior, a striking feature of
 the model of Eq.~(\ref{e.XYZhz}) is the occurrence of an 
 exactly factorized ground state for a field $h_{\rm f}(\Delta_y)$ 
 lower than the critical field $h_{\rm c}$,
 as predicted in Ref. \cite{KurmannTM82}. 
 In the case of the XYX model, this {\it factorizing field} 
 is \cite{KurmannTM82} $h_{\rm f} = \sqrt{2(1+\Delta_y)}$.
 At $h = h_{\rm f}$ the ground state of the model takes a product 
 form $|\Psi\rangle = \bigotimes_{i=1}^{N} |\psi_i\rangle $, where
 the single-spin states $|\psi_i\rangle$ are eigenstates
 of $({\bm n}_{1(2)} \cdot \hat{\bm S})$ with ${\bm n}_{1(2)}$ 
 being the local spin orientation on sublattice 1 (2).
 Taking ${\bm n} = (\cos \phi \sin\theta,
 \sin \phi \sin\theta, \cos \theta)$,
 one obtains \cite{KurmannTM82} $\phi_1 = 0$, $\phi_2 = \pi$, 
 $\theta_1 = \theta_2 = \cos^{-1}\sqrt{(1+\Delta_y)/2}$ . 
 The factorized state of the anisotropic model continuously connects 
 with the fully polarized state of the isotropic model in a field 
 for $\Delta_y =1$ and $h=2$.


 Despite its exceptional character, the occurrence of 
 a factorized state is not marked by any particular anomaly
 in the experimentally measurable thermodynamic quantities
 shown in the lower panel of Fig. \ref{f.XYXphd}. 
 However, we will now see how entanglement estimators
 are able to pin down a factorized state 
 with high accuracy. 
 
To estimate the {\it entanglement of formation} 
\cite{Bennettetal96} in the quantum 
spin chain of Eq.~(\ref{e.XYZhz}) we make use of the 
{\it one-tangle} and of the {\it concurrence}. 
The one-tangle \cite{Coffmanetal00,Amicoetal04} 
quantifies the $T=0$ entanglement of a single spin 
with the rest of the system. It is
defined as $\tau_1 = 4 \det \rho^{(1)}$, 
where $\rho^{(1)} = (I + \sum_\alpha
M^{\alpha} \sigma^{\alpha})/2$
is the one-site reduced density matrix,
$M^{\alpha} = \langle \hat{S}^{\alpha} \rangle$
(estimated as $M^{\alpha} = |\langle \hat{S}_i^{\alpha}
\hat{S}_{i+L/2}^{\alpha} \rangle|^{1/2}$
 in the QMC on a chain of length $L$), 
$\sigma^{\alpha}$ are the Pauli matrices, and $\alpha=x,y,z$.
In terms of the spin expectation values $M^{\alpha}$,
$\tau_1$ takes the simple form:
\begin{equation}
\tau_1 = 1 - 4 \sum_\alpha (M^{\alpha})^2 .
\end{equation}
It can be easily shown that the vanishing of $\tau_1$
implies a factorized ground state, and viceversa.\\
 The concurrence \cite{Wootters98} quantifies instead
the pairwise entanglement between two spins at sites
$i$, $j$ both at zero and finite temperature.
For the model of interest, in absence of
spontaneous symmetry breaking ($M^x = 0$) the 
concurrence takes the form \cite{Amicoetal04}
\begin{equation}
C_{ij}= 2~{\rm max}\{0,C_{ij}^{(1)},C_{ij}^{(2)}\}~,
\label{e.tauC}
\end{equation}
where
\begin{eqnarray}
C_{ij}^{(1)} &=&g_{ij}^{zz}-\frac{1}{4}+|g_{ij}^{xx}-g_{ij}^{yy}|~,
\label{e.C1}\\
C_{ij}^{(2)}&=&|g_{ij}^{xx}+g_{ij}^{yy}|-
\sqrt{\left(\frac{1}{4}+g_{ij}^{zz}\right)^2-(M^z)^2}~,
\label{e.C2}
\end{eqnarray}
with $g_{ij}^{\alpha\alpha}=
\langle\hat{S}_i^\alpha\hat{S}_{j}^\alpha\rangle$.
 
 \begin{figure}
 \begin{center}
\includegraphics[bbllx=0pt,bblly=-40pt,bburx=550pt,bbury=480pt,%
     width=72mm,angle=0]{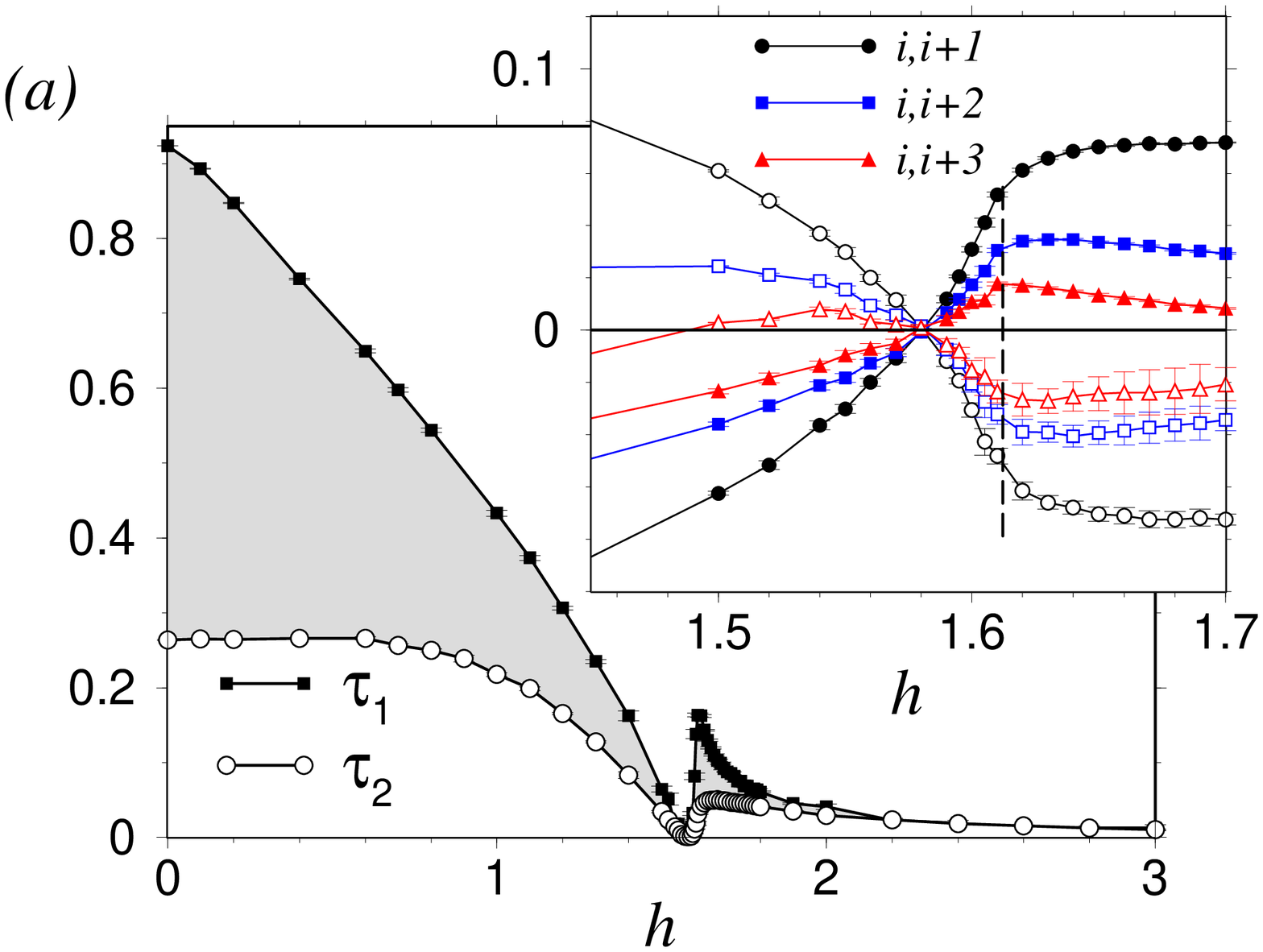}
\includegraphics[bbllx=0pt,bblly=-40pt,bburx=550pt,bbury=160pt,%
     width=72mm,angle=0]{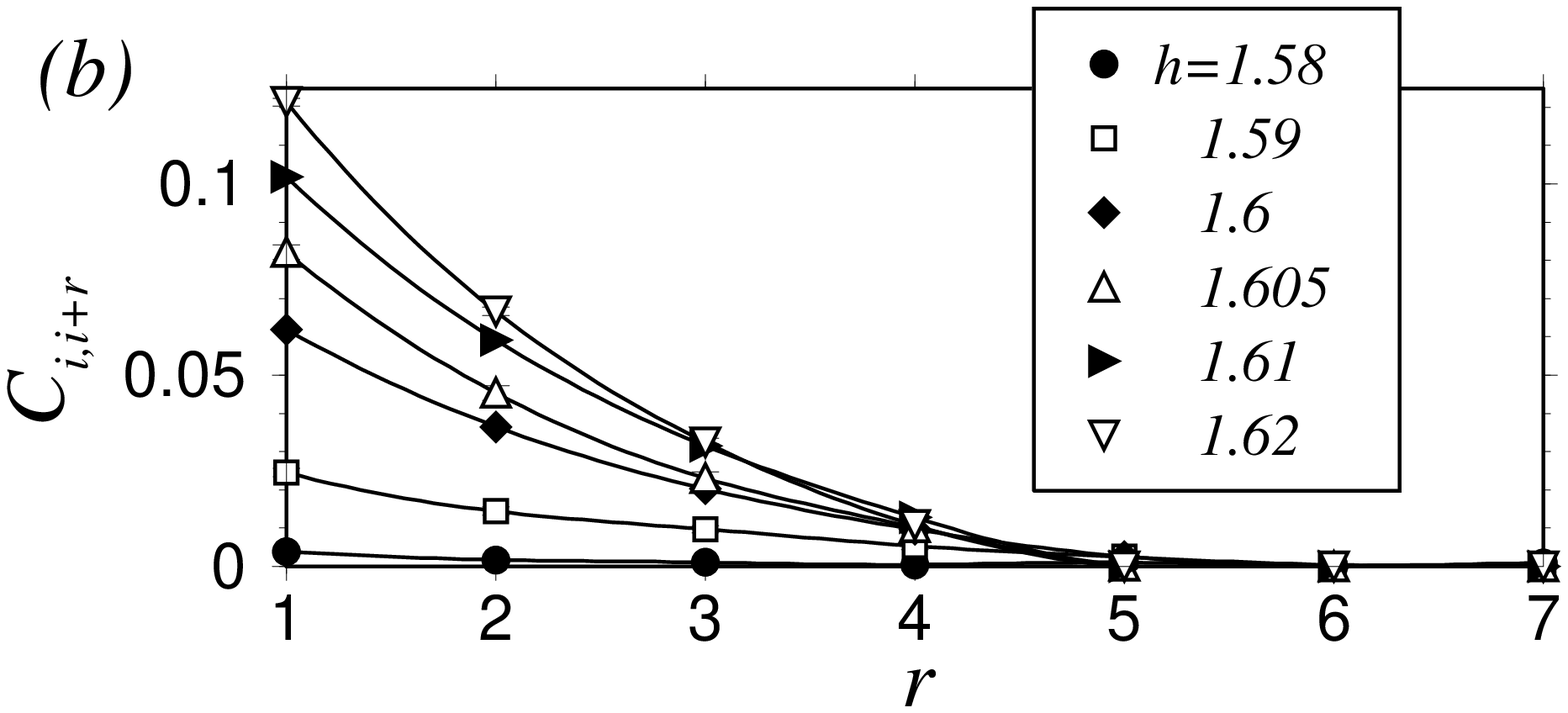}     
  \null \vskip -1cm
 \caption{\label{f.conc} (a) One-tangle $\tau_1$ and sum of squared
 concurrences $\tau_2$ as a function of the applied field 
 for the $S=1/2$ XYX model with 
 $\Delta_y = 0.25$, $L = 100$ and $\beta=200$. 
 Inset: contributions to the concurrence
 between j-th neighbors. Full symbols stand for 
 $C^{(1)}_{i,i+j}$, open symbols for $C^{(2)}_{i,i+j}$.
 The dashed line marks the critical field $h_{\rm c}$.
 (b) Concurrence as a function of spin-spin distance.
 Parameters as in the previous figure.}
 \end{center}
\end{figure}
 The $T=0$ QMC results for the model Eq.~(\ref{e.XYZhz}) with
$\Delta_y = 0.25$ are shown in Fig.~\ref{f.conc},
where we plot $\tau_1$, the sum of squared concurrences
 \begin{equation}
 \tau_2 = \sum_{j\neq i} C_{ij}^2~,
 \end{equation}
 and the field and space dependence of the concurrence.
 The following discussion, although directly referred to the
 results for $\Delta_y = 0.25$, is actually
 quite general and applies to all the other studied
 values of $\Delta_y$.  
 
 Unlike the standard magnetic observables plotted in 
 Fig. \ref{f.XYXphd}, the entanglement estimators display
 a marked anomaly at the factorizing field, where they 
 clearly vanish as expected for a factorized state. 
 When the field is increased above $h_{\rm f}$, the
 ground-state entanglement 
 has a steep recovery, accompanied by the quantum
 phase transition at $h_{\rm c} > h_{\rm f}$.
 For $\Delta_y = 0.25$, {\it e.g.}, $h_{\rm c} = 1.605(3)$ and
 $h_{\rm f} = 1.5811$.
 The system realizes therefore an interesting 
 {\it entanglement switch} effect controlled by the 
 magnetic field. 
 
 As for the concurrence, Fig. \ref{f.conc}(b) shows 
 that its range is always finite at and around the 
 critical point, and it never extends farther than the 
 fourth neighbor.
 Moreover, the factorizing field divides two 
 field regions with different expressions for the
 concurrence:
 \begin{eqnarray}
 C_{ij}^{(1)}<&0&<C_{ij}^{(2)}~~~{\rm for}~~~h<h_{\rm f}~,\\
 C_{ij}^{(2)}<&0&<C_{ij}^{(1)}~~~{\rm for}~~~h>h_{\rm f}~,
 \end{eqnarray} 
 whereas $C_{ij}^{(1)} = C_{ij}^{(2)} = 0$
 at $h=h_{\rm f}$.
   
 In presence of spontaneous symmetry breaking
 occurring for $h < h_{\rm c}$, the expression of the 
 concurrence is generally expected to change with respect
 to Eqs.~(\ref{e.C1}),(\ref{e.C2}), as extensively discussed 
 in Ref. \onlinecite{Syljuasen03} . For the model 
 under investigation, this happens when the condition 
 \cite{Syljuasen03} $C_{ij}^{(2)} < C_{ij}^{(1)}$  
 is satisfied, i.e. for $h > h_{\rm f}$. This means
 that our estimated concurrence is accurate even in the 
 ordered phase above the factorizing field; in the 
 region $0 < h < h_{\rm f}$ it represents instead a lower
 bound to the actual $T=0$ concurrence. 
 Alternatively it can be regarded as the concurrence for 
 infinitesimally small but finite temperature. 
 \begin{figure}
 \begin{center}
\includegraphics[bbllx=0pt,bblly=150pt,bburx=550pt,bbury=420pt,%
     width=85mm,angle=0]{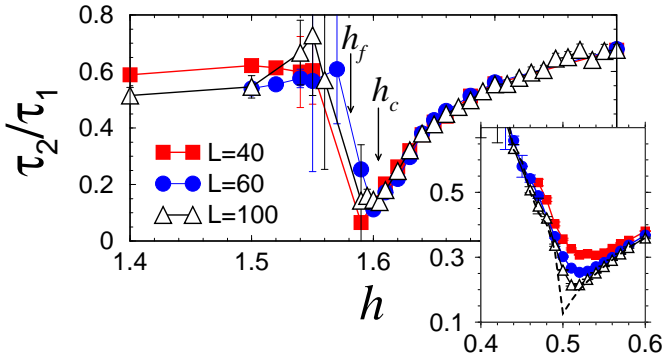}
 \caption{\label{f.ratio} Entanglement ratio $\tau_2/\tau_1$
 as a function of the field for $\Delta_y = 0.25$
 and $\beta=2L$.
 Inset: entanglement ratio for the 1D Ising model in a transverse
field. The dashed line is the $L\to\infty$ extrapolation result.} 
 \null \vskip -1cm
 \end{center}
\end{figure} 

 In Fig. \ref{f.conc} the sum of squared concurrences $\tau_2$ 
 is always smaller than or equal to the one-tangle $\tau_1$, in 
 agreement with the Coffman-Kundu-Wootters conjecture
 \cite{Coffmanetal00}. This shows that entanglement is 
 only partially stored in two-spin correlations, and it is
 present also at the level of three-spin entanglement, 
 four-spin entanglement, etc. 
 In particular, we interpret the {\it entanglement ratio}
 $R = \tau_2/\tau_1$ as a measure of the fraction of the total 
 entanglement stored in {\it pairwise} correlations. This
 ratio is plotted as a function of the field in Fig. \ref{f.ratio}.
 As the field increases, we observe the general trend 
 of pairwise entanglement saturating the whole entanglement
 content of the system. But a striking anomaly occurs at 
 the quantum critical field $h_c$, where $R$ displays a very 
 narrow dip. According to our interpretation, this result
 shows that the weight of pairwise entanglement decreases
 dramatically at the quantum critical point in favour of 
 {\it multi-spin entanglement}. Unlike classical correlations, 
 entanglement shows the special property of {\it monogamy}
 \cite{Coffmanetal00}, namely full entanglement
 between two partners implies the absence of entanglement with the 
 rest of the system. Therefore multi-spin entanglement 
 appears as the only possible quantum counterpart
 to long-range spin-spin correlations occurring at a quantum
 phase transition.  
 This also explains the somewhat puzzling
 result that the concurrence remains short-ranged at a quantum
 phase transition (Fig. \ref{f.conc}(b)) while the spin-spin 
 correlators become long-ranged, and it evidences the 
 serious limitations of concurrence
 as an estimate of entanglement at a quantum critical point. 
Strong indications on the relevance of 
multi-spin entanglement in quantum-critical spin chains come also 
from the study of the entanglement between a block of $L$ contiguous 
spins and the rest of the chain\cite{Vidaletal03}. 
Finally, multi-spin entanglement involves a macroscopically 
 coherent superposition of quantum states, and this result is
 consistent with the picture of macroscopic ({\it i.e.}
 long-wavelength) quantum fluctuations occurring at a quantum 
 phase transition. 
 
 In turn, we propose the minimum of the entanglement 
 ratio $R$ as a novel estimator of the quantum critical
 point, fully based on entanglement quantifiers. 
 This result appears general for the whole class of 
 models described by the hamiltonian of Eq.~(\ref{e.XYZhz}).
 Inset (b) of Fig. \ref{f.ratio} shows in fact that an
 analogous dip in the entanglement ratio signals the quantum
 phase transition in the Ising model in a transverse field
 ($\Delta_y=\Delta_z=0$), occurring at the critical field 
 $h_c=1/2$. Work is in progress to test the universality of 
 such novel signature of quantum critical behavior for
 completely independent quantum phase transitions.
 
\begin{figure}
\begin{center}
\null \hspace{-3cm}
\includegraphics[
bbllx=0pt,bblly=250pt,bburx=450pt,bbury=480pt,%
     width=65mm]{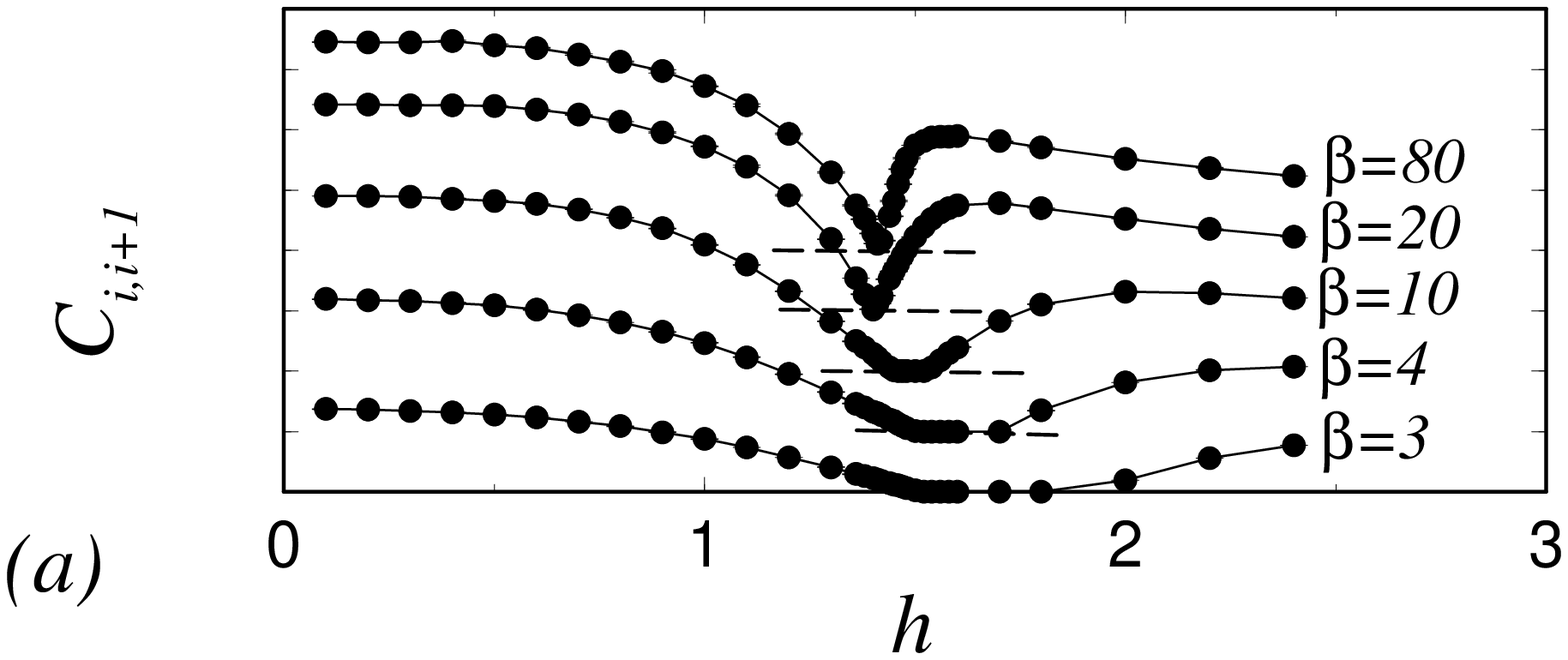}
 \includegraphics[
bbllx=100pt,bblly=180pt,bburx=500pt,bbury=500pt,%
     width=60mm]{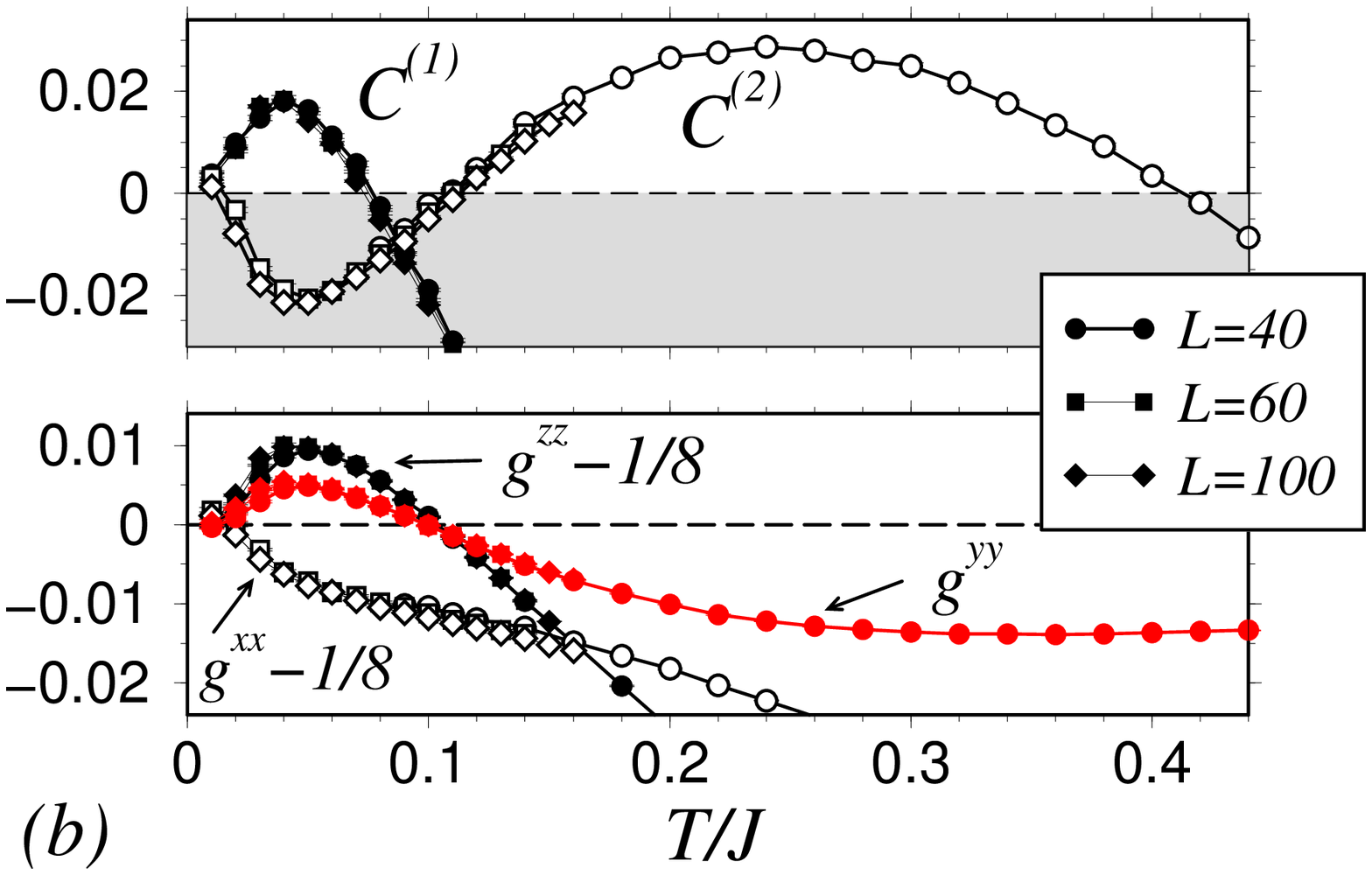}    
 \caption{\label{f.thermalC} (a) Nearest-neighbor concurrence
 $C_{i,i+1}$ as a function of field for different temperatures
 in the XYX model with $\Delta_y = 0$, $L=40$. 
 One division on the y axis corresponds to a concurrence interval
 of 0.1. Offsets (marked by the dashed lines) have been 
 introduced to improve readability. (b) Components of the 
 nearest-neighbor concurrence and spin-spin correlators as
 a function of temperature
 at the factorizing field $h_{\rm f}=\sqrt{2}$. The dashed
 lines mark the zero value.}
 \null \vskip -0.5cm
 \end{center}
\end{figure}

  The use of the QMC method enables us
 to naturally monitor the fate of entanglement
 when the temperature is raised above zero. In this regime
 the concurrence is the only well-defined
 estimator of entanglement, whereas the one-tangle has not
 yet received a finite-temperature generalization. 
 Fig. \ref{f.thermalC}(a) shows the nearest-neighbor (n.n.) 
 concurrence as a function of the field for different temperatures 
 in the XYX model
 with $\Delta_y = 0$. We observe that $C_{i,i+n} < C_{i,i+1}$
for $n>1$, and at high enough temperature ($T\gtrsim 0.1J$)
only the n.n. concurrence survives. The most prominent feature is
 the persistence of a field value (or an interval of values) 
 at which the concurrence is either zero (for $T \gtrsim 0.05J$) 
 or $\sim 10^{-3}$ (for $0 < T \lesssim 0.05J$). 
 In particular, the field values for which the concurrence
 vanishes are temperature-dependent, so that two-spin 
 entanglement can be switched on
and off tuning both the field and the temperature. 
%
%
 Fig. \ref{f.thermalC}(b) shows a highly
 non-trivial temperature dependence of the two-spin entanglement
 at $h=h_f = \sqrt{2}$. Although vanishing
 at $T=0$, the n.n. concurrence has a quick thermal activation 
 due to thermal mixing of the factorized ground state with
 entangled excited states. Although the spectrum over the ground
 state displays a gap of order $0.1 J$ \cite{Dmitrievetal02},
 in one dimension strong fluctuations induce {\it thermal 
 entanglement} \cite{Arnesenetal01} already at temperatures
 which are an order of magnitude lower. The appearence 
 of thermal entanglement is directly related to the 
 {\it increasing} behavior of the correlators $g^{yy}=g^{yy}_{i,i+1}$
 and $g^{zz}=g^{zz}_{i,i+1}$ entering the expression of the $C^{(1)}$
 component, Eq. (\ref{e.C1}) (also shown in Fig. \ref{f.thermalC}(b)). 
 In particular the appearence of a finite  $g^{yy}$ is a purely 
 quantum effect, since $\Delta_y=0$.  
 Because of the non-monotonous behavior of the 
 correlators, at higher temperatures thermal entanglement disappears and 
 reappears again, revealing an intermediate temperature region where 
 two-spin entanglement is absent.


  In summary, making use of efficient QMC 
 techniques we have provided a comprehensive picture of
 the entanglement properties in a class of anisotropic
 spin chains of relevance to experimental compounds. 
 We have shown that the occurrence of a classical 
 factorized state in these systems is remarkably singled
 out by entanglement estimators, unlike the more
 conventional magnetic observables. Moreover we find
 that entanglement estimators are able to detect 
 the quantum critical point, marked by a narrow dip in 
 the pairwise-to-global entanglement ratio.
 Therefore we 
 have shown that entanglement estimators provide 
 precious insight in the ground-state properties of
 lattice $S=1/2$ spin systems. Thanks to the versatility
 of QMC, the same approach can be used
 for higher-dimensional systems. In this respect,
 investigations of the occurrence of factorized states
 in more than one dimension are currently in progress. 
 Finally, the proximity of a
 quantum critical point to the factorized state of the
 system gives rise to an interesting field-driven 
 entanglement-switch effect. This demonstrates that many-body 
 effects driven by a macroscopic field are a  
 powerful tool for the control of the microscopic
 entanglement in a multi-qubit system, and stand as 
 a profitable resource for quantum computing devices.   
 
 Fruitful discussions with L. Amico, T. Brun, P. Delsing, 
G. Falci, R. Fazio, A. Osterloh, and G. Vidal
are gratefully acknowledged. We acknowledge support
by NSF under grant DMR-0089882 (T.R. and S.H.),
by INFN, INFM, and MIUR-COFIN2002 (A.F., P.V., and V.T.).


\begin{thebibliography}{200}
\bibitem{note} The original directed-loop algorithm
 was designed to treat models with a continuous
 rotational symmetry in the plane transverse to the field.
 We have generalized it to our less symmetric model 
 introducing further vertices not conserving the $z$-magnetization, 
 associated with the operators $S_i^{+}S_{i+1}^{+}$ and 
 $S_i^{-}S_{i+1}^{-}$ appearing in our hamiltonian with 
 quantization axis along $z$. We then minimize the bounce
 probability as in the original scheme of Ref. \onlinecite{SyljuasenS02}.
 Details of the algorithm will be presented elsewhere.
\end{thebibliography}
\end{document}